\documentstyle[12pt,aasms4]{article}
\def \cf           {{cf. }}

\def \eg           {{e.g.}}
\def \etal         {{et~al. }}
\def \ie           {{i.e.}}

\def \kms          {\hbox{km$\,$s$^{-1}$}}

\def\approxlt{\lower.2em\hbox{$\buildrel < \over \sim$}}
\def\approxgt{\lower.2em\hbox{$\buildrel > \over \sim$}}

\def \ls           {\hbox{L$_{\odot}$}}
\def \ms           {\hbox{M$_{\odot}$}}           % Solar mass
\def \ni           {\noindent}
\def \psec         {$.\negthinspace^{\prime\prime}$}
 
\def \date         {\ifcase\month \message{zero} \or
                    January \or February \or March \or April \or May \or June 
                    \or July \or 
                    August \or September \or October \or November \or 
                    December \fi
                    \space\number\day, \number\year}

\begin{document}
\title{Arp 302: Non-starburst Luminous Infrared Galaxies}

\author{K.Y. Lo, Yu Gao, and Robert A. Gruendl}
 
\affil{Laboratory for Astronomical Imaging, \\
Department of Astronomy, University of Illinois, \\
    1002 West Green Street, Urbana, IL 61801}

\authoremail{kyl, gao, gruendl@astro.uiuc.edu}

%\received{\date}
 
\begin{abstract}
Arp 302, a luminous infrared source 
($L_{\rm IR} = 4.2 \times 10^{11} ~\ls$), 
consisting of two spiral galaxies (VV340A and VV340B) with nuclear 
separation of $\sim 40''$, has the highest CO 
luminosity known. Observations with the BIMA array at 
5\arcsec\ $\times $7\arcsec\ resolution reveal that the CO emission
is extended over 23.0 kpc in the edge-on spiral galaxy, VV340A,
corresponding to $6.7 \times 10^{10}~\ms$ of H$_2$.
In the companion face-on galaxy, VV340B, the CO emission
is extended over $\sim$ 10.0 kpc, with $1.1 \times 10^{10}~\ms$ of H$_2$. 
The large CO extent is in strong contrast to starburst systems, such
as Arp 220, in which the CO extent is typically $\le$ 1 kpc. 
Furthermore, $L_{\rm IR}/M$(H$_2)$ is found to be 
$\approxlt$ 6.0 ~\ls/\ms ~throughout both galaxies.
Thus the high IR luminosity of Arp 302 is apparently 
not due to starbursts in the nuclear regions, but is due to its 
unusually large amount of molecular gas forming stars at a rate
similar to giant molecular clouds in the Milky Way disk. Arp 302 
consists of a pair of very gas-rich spiral galaxies that may be 
interacting and in a phase before a likely onset of starbursts. 

\end{abstract}
 
\keywords{galaxies: individual (Arp 302, VV340, UGC 09618) -- 
galaxies: interactions -- galaxies: spiral -- galaxies: starburst 
-- infrared: galaxies -- ISM: molecules}
 
\section{INTRODUCTION}

Ultra-luminous infrared (IR) galaxies ($L_{\rm IR} \ge 10^{12} ~\ls$) 
and most luminous IR galaxies (LIRGs; $10^{12} ~\ls > 
L_{\rm IR} \ge 10^{11} ~\ls$) are almost always found in 
interacting or merging systems (\eg, Sanders \etal 1988; 
Melnick \& Mirabel 1990; Murphy \etal 1996).  On the other hand, 
only a small fraction of optically selected interacting/merging 
galaxy systems are truly IR-luminous (\eg, Arp 1966; 
Arp \& Madore 1977; Kennicutt \etal 1987; Bushouse \etal 1988). 
Apparently, galaxy interaction is a common causal factor for 
the enhanced IR luminosity, but it is not the only one 
(\cf Lawrence \etal 1989; Leech \etal 1994). 
 
As the large IR luminosity can be explained by a burst 
of star formation and almost all LIRGs are found to be molecular 
gas-rich (Sanders \etal 1991; Solomon \etal 1996),
understanding of LIRG mergers then depends 
on how interaction induces starbursts and enhances the star formation
efficiency in these merging systems. A study of the gas properties
at various phases of the merging process would help
identify the key physical processes involved.  For example,
all four ultra-luminous IR galaxies and most LIRGs in advanced 
merging stages observed with sufficiently high spatial resolution, show 
a high CO surface brightness in their nuclear region 
with an extent $\la$ 1 kpc typically.  This has been
interpreted as a very high central H$_2$ surface density, usually more
than 10$^4$ ~\ms~${\rm pc^{-2}}$ (Scoville \etal 1991; Yun \& Scoville 1995).

Few studies have been able to explore the gas-rich galaxy 
mergers over various merging phases (\eg, Combes \etal 1994). In a 
study of 50 LIRG mergers, the CO luminosity is found to
increase with increasing separation of the merging nuclei, which may
imply that the molecular gas content of LIRGs is decreasing as 
 merging advances (Gao 1996; Gao \& Solomon 1996). 
Previous studies have concentrated on relatively more advanced merger
systems in which the interstellar medium (ISM) has already been highly 
disrupted by the starbursts.  In order to isolate the conditions in the ISM
{\it leading} to starbursts, we are starting a program to study a 
sample of LIRGs
chosen to represent different phases of the interacting/merging 
process, using the newly expanded 
Berkeley-Illinois-Maryland Association (BIMA) millimeter-wave array 
(Welch \etal 1996) which is ideally suited to this 
problem given its large primary beam and wide spectral bandwidth.
The goal is to sample statistically the evolution of the properties of 
the molecular
clouds in the galaxies as they go through the merging processes.

Here, we present the first results of this program on Arp 302.
Arp 302 (UGC 09618) consists of two galaxies (VV340A and VV340B 
-- Sb and Sc, 
with apparent magnitudes 14.6$^{\rm m}$ and 15.3$^{\rm m}$, 
respectively) with a separation of $\sim 40''$
between the two nuclei. At a median redshift (cz) of \ 10160 \kms, \ its
luminosity distance, \ d$_{\rm L}$, \ is 136.6 Mpc ($H_0=75$ \kms Mpc$^{-1}$,
q$_0$= 1/2) and it has an infrared luminosity, \ $L_{\rm IR}$, \ of \ 
$4.2 \times 10^{11} \ \ls$.  

 Single-dish measurements show a CO 
luminosity $L_{\rm CO} \sim 1.9 \times 10^{10}$ K~\kms ~pc$^2$, 
which corresponds to a molecular gas mass of $\sim 9\times 10^{10} ~\ms$, 
the largest amount of molecular gas known among the samples of $\sim 50$ 
LIRG mergers studied recently by Gao (1996) and $\sim 37$ ultra-luminous 
galaxies surveyed by Solomon \etal (1996).  Given the large IR luminosity
and the close proximity of the two galaxies, the normal assumption would 
be that Arp 302 is powered by starbursts that
result from the interaction of the two galaxies.
The relative large separation ($\sim$ 25 kpc = $\sim 40''$ at the 
angular distance of 136(1+z)$^{-2}$ = 128 Mpc) compared 
to merger remnants such as Arp 220 would further suggest that Arp 302 is in
an early phase of merging/interacting.

\section{OBSERVATIONS AND DATA REDUCTION}

Observations of Arp 302 were started with the BIMA 6-element array 
in October 1995. More observations were made
with the new 9-element array in the C configuration in April and B 
configuration in May 1996 (Welch \etal 1996).  The digital 
correlator was used with the widest bandwidth to cover a velocity 
range of 1,600 \kms, at a resolution of 8.4 \kms.  The  
$\sim 2'$ primary beam was centered between the two galaxies to cover 
all possible emission in this interacting system which extends slightly
more than $1'$. The observing frequency was 111.4952 GHz, corresponding to 
the CO(1-0) line at the median redshift cz=10,160 \kms \ of this 
merging galaxy system.

The nearby quasar 3C345, used for the complex antenna gain 
calibration, was observed for 8 minutes before every 30 minutes on
Arp 302. The strong quasar 3C273 was observed at the beginning
for 30 minutes to calibrate the spectral dependence of the gain. 
The typical system temperature ranged between 300 and 800 K. 

The interferometer data were reduced with the MIRIAD data reduction 
package (Sault \etal 1995).  Observations in May 1996 on baselines longer than 
50 k$\lambda$ were found to suffer from atmospheric decorrelation
and were therefore weighted by applying a Gaussian taper (effective
resolution of 3\psec 5) in the UV plane prior to inverting the UV data cube.
Data cubes of the CO brightness distribution, ${\rm T_b(x,y,V)}$, 
where x and y are the spatial coordinates and V is the radial velocity,
at various velocity 
resolutions (8.4, 20.0 and 50.0 \kms) were created covering
the range from $-$600 to 600 ~\kms ~about cz = 10,160 \kms. 
The cleaned maps have a final 
synthesized beam of $7''\times 5''$, with the major axis of the
beam at a position angle of 10$^{\circ}$. The various moment maps were 
then made to produce the integrated intensity map 
(${\rm I_{CO}(x,y) = \int  T_b(x,y,V) dV}$  in K-\kms), 
the intensity weighted velocity map, and the intensity weighted velocity 
dispersion map.  

\section{RESULTS}

\subsection{Spatial Distribution of the CO Emission}

In Figure 1 (Plate), the contours of an integrated CO intensity map
at $7''\times 5''$ resolution  are superposed on a R-band CCD image 
of Arp 302 (left panel); 
whereas the right panels show the CO spectra from various locations
in Arp 302.   The $\rm I_{CO}$ distribution is very well correlated 
with the R-band CCD image of the galaxies, especially along the
prominent dust lane in the northern edge-on galaxy VV340A. The CO
emission is especially bright in the inner $\sim 14''$, or $\sim$8.7 kpc.  
At the angular distance of 128 Mpc, $1''$ corresponds to 620 pc.

About $\sim 85\%$ of the total CO flux
(${\rm S_{CO} = (2k/\lambda^2)~\int T_b(x,y,V) dxdydV}$ in Jy~\kms) 
is from the 
edge-on galaxy, distributed along the major-axis
over $\sim 35''$ and marginally resolved along the minor axis.
If the CO emission
is coplanar with the galactic disk, the shape of the CO
distribution would place a limit on the inclination angle, i,
of the galaxy to be $\approxgt 75^\circ$.

While $\sim 15\%$ of the total CO flux is from the southern 
galaxy VV340B, the CO surface brightness is weaker, well resolved by 
the $7''\times 5''$ beam, and appears correlated with the central bar.   
Some correspondence of weak CO emission with the northern spiral arm
is also indicated (Fig. 1).  Within the uncertainties, all the CO 
flux detected from these two galaxies with the NRAO 12m telescope 
(Gao 1996) is detected in the BIMA maps. 

We have also detected CO emission from a position away from the two
galaxies at the position: $\alpha$(J2000.0)=14$^h$57$^m$01.$\negthinspace^s$6, 
$\delta$(J2000.0)=24$^\circ$36\arcmin 48\arcsec ~(\cf spectrum labelled Off
in Fig. 1) that does
not correspond to any optical feature.  The reality of this interesting
source has to be verified by further observations.

\subsection{Kinematics}

Figure 2 shows a position-velocity diagram of the CO surface
brightness in VV340A along the major 
axis (PA $\sim 5^\circ$) with $\sim$6\arcsec\ and 20 \kms\ resolution.
It shows that the kinematics of the molecular gas in VV340A
has largely a constant velocity gradient (85.5 \kms ~kpc$^{-1}$) 
within the central $14''$, or $\sim$8.7 kpc, along the major axis
and a velocity range of 780 \kms\ at FWZI.  The faint CO emission 
to the north of the central 15\arcsec\ ($\sim 10$ kpc)
appears to have 
a constant velocity between 260 \kms\ and 344 \kms ,
and to the south between $-$180 \kms\ and $-$210 \kms .
At the extreme southern end of the disk, the integrated intensity map (Fig. 1)
shows a slight deviation westward from the linear morphology exhibited by
the rest of the galaxy.  In the same region, the velocity field also has
a distortion from the rotational signature.  This could be interpreted 
as evidence that the interaction between the two galaxies has perturbed 
the molecular gas distribution.  

A constant velocity gradient has often been described as indicating
a ``solid-body'' rotating disk.  In the case of the inner $14''$ of VV340A,
the ``solid-body'' disk would have a 4.35 kpc radius.  Alternately, a 8.7 kpc 
diameter rotating molecular ring, while also consistent with the constant 
velocity gradient, would be easier to understand dynamically.
A ring of molecular gas may also explain 
the shape of the CO spectrum from VV340 A (Fig. 1), and
the weakness of CO emission near the systemic velocity
($\sim 40$ \kms in the figures, or cz = 10200 \kms) from the nuclear region.
Thus, a strong concentration of molecular gas in a ring structure 
rotating with a circular speed of 390 \kms\ (corresponding to the inner disk)
combined with a thin, extended, molecular gas disk with a flat 
rotation velocity of $\sim$240 \kms\ viewed edge on would be consistent with 
the observations.  We note, however, that the observed
systemic velocities of the 
rotating ring and disk, if taken as the centroid of the extent of each 
respective component, differ by $\sim$30 \kms .  This may arise if the 
distribution of material in the plane of VV340A is not axi-symmetric and
has non-circular motions. Further study of the gas kinematics is 
necessary.  

There is a velocity gradient across VV340B as well, although
the total velocity extent across VV340B is only $\sim 100$ \kms
(FWZI).  Given the large amount of molecular gas and the narrow
line-width, the galaxy must be nearly face-on, which is 
consistent with the shape of the galaxy in the R--band image (Fig. 1).

From the centroid of the CO line profiles for VV340A and B,
we determine the relative systemic velocity between the two galaxies,
${\rm V_A - V_B = -340 \pm 20 ~\kms}$, with VV340B blueshifted
relative to VV340A.  Thus, the spin vector of VV340A could be nearly 
parallel to the orbital angular momentum vector of VV340B, unless a very
large relative transverse motion exists.

\subsection{Molecular Mass and Surface Density}

Arp 302 is the most CO luminous system known in the local universe, 
more luminous in CO than those ultra-luminous galaxies 
out to the redshift of $\sim 0.27$ (Solomon \etal 1996).  
The total molecular gas mass of each 
galaxy can be estimated from the total CO flux,
using ${\rm M({\rm H}_2)=1.18\times 10^4 d_L\negthinspace^2 S_{CO}}$, 
where ${\rm d_L}$ 
is the luminosity distance in Mpc and ${\rm S_{CO}}$ is the total CO flux 
in Jy \kms.  We use the conversion factor, ${\rm X \equiv [N(H_2)/I_{CO}] =
3.0 \times 10^{20} ~cm^{-2}/(K \kms)}$ for convenient comparison to previous
results, even though it is not a priori applicable to Arp 302.  

The H$_2$ gas mass in VV340A is $(6.7 \pm 0.3) \times 10^{10} ~\ms$,
with \approxgt 85\% in the central 8.7 kpc.  The dynamical mass interior 
to the rotating molecular ring (4.35 kpc radius) is estimated to be 
$\sim 1.5 \times 10^{11} ~\ms$. This implies that if He is included, 
the gas mass constitutes \approxgt 60\% of the dynamical 
mass of VV340A within a radius of 4.35 kpc.  The H$_2$ gas mass in VV340B 
is $(1.1 \pm 0.1) \times 10^{10} ~\ms$.  The combined H$_2$ gas mass in 
this merging system is ($7.9 \pm 0.3) \times 10^{10} ~\ms$.  

The HI mass in Arp 302 has been measured to be 
\approxgt $2.3 \times 10^{10} ~\ms$ (Mirabel \& Sanders 1988).  
Thus, the total gas (HI + H$_2$ + He) content of Arp 302 is 
$\sim 1.4 \times 10^{11} ~\ms$.  Even if X is 5 times
smaller, the total gas content would still be $\sim 0.54 \times 10^{11} ~\ms$.
However, this does point to the importance of a direct determination of 
the conversion factor X for Arp 302.

Although the southern end of the VV340A molecular disk is distorted, 
no molecular gas has been detected between 
the two galaxies in the disk overlap region.  But, the CO source to the
east of VV340A has a molecular mass of $(1.0 \pm 0.2) \times 10^9$ \ms.

The spatial extent of the CO source in Arp 302 provides
estimates of the face-on ${\rm H_2}$ surface density, 
$\Sigma_{\rm H_2}$.  Thus, we obtain $\Sigma_{\rm H_2}$
of ${\rm \sim 10^3 ~\ms~pc^{-2}}$ within the radius 
of 4.35 kpc and ${\rm \sim 20 ~\ms~pc^{-2}}$ in the outer
disk in VV340A, and ${\rm \sim 100 ~\ms ~pc^{-2}}$ for the
inner part of VV340B.  In the inner parts of both galaxies,  
$\Sigma_{\rm H_2}$ is high compared to that of the Milky Way disk
(4 $-$ 17 \ms~${\rm pc^{-2}}$; Sanders \etal 1984),
but small compared to that of the LIRGs and ultra-luminous galaxies 
at the more advanced
merger stage (typically $> 10^4 ~\ms ~{\rm pc^{-2}}$; 
\eg, Yun \& Scoville 1995).

\section{DISCUSSION}

\subsection{Is Arp 302 a Star-burst System ?}

While the $L_{\rm IR} = 4.2 \times 10^{11} ~\ls$ certainly qualifies
Arp 302 as a LIRG, the CO emission is extended over 
the entire $\sim$23 kpc diameter disk in VV340A and over
many kpc in VV340B.  The large CO extent 
is in stark contrast to most of the more than 20 LIRGs and 
ultra-luminous galaxies
for which high spatial resolution CO(1-0) observations 
have been made.  They showed highly concentrated CO emission
in nuclear regions $\la 1$ kpc in size typically (Scoville \etal 1991). 

Spatially unresolved observations of LIRGs showed that they
tend to have very large $L_{\rm IR}$ \ and \ $L_{\rm CO}$, usually 
interpreted as indicating very high star formation rate 
and H$_2$ gas mass, as well as very high star formation efficiency
(\ie, $L_{\rm IR}/M$(H$_2)$; Solomon \& Sage 1988; Young \etal 1989;
Sanders \etal 1991; 
Solomon \etal 1996).  From spatially resolved observations, one can
derive local measures of CO surface brightness both in terms of
${\rm T_b}$ and ${\rm I_{CO}}$ converted to $\Sigma_{\rm H_2}$,
the H$_2$ surface density. For LIRGs and ultra-luminous galaxies, 
all these quantities
are extremely high compared to Galactic values (\eg, Scoville \etal 1991;
Sargent \& Scoville 1991; Yun \etal 1994; Yun \& Scoville 1995)

The ratio of $L_{\rm IR}/M({\rm H_2)}$, used to indicate
the star formation efficiency,
ranges between 20 to 100 ~$\ls/\ms$ for nuclear
starburst regions (Scoville \etal 1991).  For Arp 302, the global
ratio is 5 $\ls/\ms$.  If we scale the IR luminosity and extent
with those of the radio continuum emission from Arp 302 (Condon \etal
1990; Condon \etal 1991), we obtain values of 6,
2.5 and 2 $\ls/\ms$
for $L_{\rm IR}/M$(H$_2)$ in the inner 8.7 kpc region and the outer part of 
VV340A, and VV340B, respectively.  This is comparable to the
ratio for giant molecular clouds (GMCs) in the Milky Way 
($L_{\rm IR}/M$(H$_2) \sim 4$, Scoville \&
Good 1989) but much 
below that observed in Arp 220 (95 $\ls/\ms$) and other starburst systems
(\cf Scoville \etal 1991). 
Furthermore, we can infer a mean IR surface brightness ($\Sigma_{\rm IR}$) ~of 
7000, 55 and 300 ~$\ls ~{\rm pc^{-2}}$
for the same three regions respectively.  This is very low compared to
the typical $\sim 10^5 ~\ls ~{\rm pc^{-2}}$ in nuclear starburst regions
(Lo \etal 1987, and \cf Scoville \etal 1991).  

Therefore, the high IR luminosity in Arp 302 arises simply because 
an extremely large amount of molecular gas is forming high mass stars 
at a rate similar to the GMC's in the disk of the Milky Way and not 
because starbursts are occurring in the galactic nuclei.

\subsection{Is Arp 302 an Interacting System ?}

The two galaxies (VV340A and VV340B) in Arp 302 have a projected 
nuclear separation of $40''$
and a velocity difference of 340 \kms. Are the two galaxies physically
associated, and not just chance superposition ?  It has been shown via
Monte Carlo simulations
that the posterior probability of two galaxies brighter than 15$^{\rm m}$
with a velocity difference of between 200 and 1000 \kms ~being bound is
0.88 (Schweizer 1987). Furthermore, the distortion of the CO distribution 
in the southern end of 
VV340A may be an indication that the two galaxies are in fact interacting.
The HI emission line profile from Arp 302 
 (U 09618 in Mirabel \& Sanders 1988) is
 significantly different from the CO profile, in that HI emission
 appears to be missing at the velocities corresponding to VV340A.
 Future VLA observations of HI in Arp 302 may show direct evidence for 
 the interaction of the two galaxies in terms of HI tidal tails and 
the kinematic signature.
 
We can also estimate the ratio of the kinetic and potential
energies of the interacting system: 
${\rm T/U = (M_r V^2/2)/(GM_A M_B/R)}$, where ${\rm M_r}$ is the reduced
mass, V is the relative velocity, and R is the nuclear separation of
the two galaxies with masses, ${\rm M_A \ and \ M_B}$.  If  
${\rm M_A \approx M_B}$ as implied from their comparable extent,\  
${\rm M_r \approx M_B/2}$, \ and \ ${\rm T/U = RV^2/(4GM_A)}$. 
For ${\rm M_A = R_AV_r^2/G}$, where V$_{\rm r}$ is the rotational velocity
of VV340A at the outermost radius R$_{\rm A}$, 
${\rm T/U = (V/V_r)^2(R/R_A)/4}$.  Since ${\rm V_r \approx 250 ~\kms}$ 
with V \approxgt 340 \kms
~(if the transverse motion is taken into account) \ and \ R $\ga 2\times 
{\rm R_A}$ (the projected separation), it is plausible that
T/U is on the order of unity and E = T $-$ U $\sim 0$.
(If the mass of one galaxy is substantially less than the other,
the T/U ratio would change by a factor of two.)  This analysis suggests
that the two galaxies could be on parabolic orbits about each other.

The relative radial systemic velocity of the two galaxies is 
340 \kms ~and the spin of VV340A is 
parallel to the apparent orbital motion. This  indicates
 that the disk of VV340A may be undergoing a prograde 
encounter with its perturber VV340B,
unless there is a very large transverse motion between 
the two galaxies.  Numerical simulations have shown that 
two gas rich galaxies of comparable mass released on  
parabolic orbits, with the mass and orientations similar to those
found in Arp 302, would merge within two orbital times,
or a few times 10$^8$ years (Barnes and Hernquist 1991; see also
White 1979; Toomre \& Toomre 1972). 

Thus, we conclude that the high probability that the two galaxies
in Arp 302 are bound, the distortion in the CO distribution in VV340A,
the plausibly small total energy of the two galaxies, and the numerical 
simulation results are all indications to suggest 
that Arp 302 is an interacting system.

Given both the stars 
and the gas distribution in Arp 302 system do not appear highly disturbed, 
the two galaxies are apparently approaching each other 
for the first time, suggesting that Arp 302 is in
an early phase of the interaction.  The next phase of 
Arp 302 may be
represented by the galaxy VV114 (Yun \etal 1994), in which the
molecular gas has merged into a common envelop
before turning into an ultra-luminous galaxy.

\section{SUMMARY}

With the BIMA array, we have resolved the CO emission from the luminous IR
galaxy system Arp 302 into galactic scale distributions.  Local measures 
in Arp 302,
such as $\Sigma_{\rm H_2}, L_{\rm IR}/M({\rm H_2}), \Sigma_{\rm IR}$,
are all much smaller than those found in starburst systems.  Thus,
the enormous IR luminosity of Arp 302 is apparently not due to starbursts,
but is due to its unusually large amount of molecular gas forming stars 
at a rate similar to giant molecular clouds in the Milky Way disk.

Also, evidence suggests that the two galaxies in Arp 302 are interacting
and at an early phase of the interaction.  As the interaction progresses,
bursts of star formation may eventually be initiated.  If so, we may 
be witnessing two interacting gas-rich galaxies in a pre-starburst phase.
Identifying and studying 
such systems are important for isolating the conditions in the ISM that 
will {\it lead} to starbursts.  

{\it Acknowledgment:} The writing of this paper was started when KYL 
was visiting the Max Planck Institut f\"ur Extraterrestrische physik
during May of 1996 under an Alexander von Humboldt Research Award. 
KYL wishes to thank R. Genzel for his hospitality and valuable 
discussions during the visit and E. Skillman for
valuable inputs.  YG thanks P.M. Solomon for discussions and 
advice during the early stages of this project.   
We thank Drs J. Mazzarella, D.C Kim
(IPAC, Caltech) and D.B. Sanders for allowing us to use the R-band 
image of Arp 302 before publication.   We are also very grateful to one
of the referees for pointing out an embarrassing error in an earlier 
version of the paper. Research
at the Laboratory of Astronomical Imaging is funded by NSF grant 93-20239
and by the University of Illinois.

\begin{figure}
 
\caption{Integrated CO(1--0) intensity (I$_{\rm CO}$) contours are 
overlaid on a R-band image (grayscale) of Arp 302 (VV340A and VV340B). 
The contours shown are 
6, 8, 12, 24, 48, and 96 $\times ~3.2 $ K-\kms\, with a
$7''\times 5''$ beam.  To the right of the image, the 
top and bottom spectra, derived from the data cube,
 show the flux density (S$_\nu$ in Jy)
of the CO emission from the two entire galaxies VV304A 
and VV304B, respectively.  The four spectra shown between the 
top and bottom spectra are the flux density of the CO emission
from a single beam area centered as follows:
North ($\alpha$=14$^h$57$^m$0.$\negthinspace^s$8, 
$\delta$=24$^\circ$37\arcmin 20\arcsec );
South ($\alpha$=14$^h$57$^m$0.$\negthinspace^s$55, 
$\delta$=24$^\circ$36\arcmin 48\arcsec );
Off  ($\alpha$=14$^h$57$^m$1.$\negthinspace^s$65, 
$\delta$=24$^\circ$36\arcmin 48\arcsec) ;
and Arm ($\alpha$=14$^h$57$^m$0.$\negthinspace^s$2, 
$\delta$=24$^\circ$36\arcmin 34\arcsec ). 
0 \kms ~corresponds to cz = 10,160 \kms.}
\end{figure}

\begin{figure}
\caption{Position--Velocity diagrams for the two galaxies VV304A (a) 
and VV304B (b) in Arp 302 system.
The contours for both panels represent T${\rm _b}$, starting at 0.064 K
and increasing by steps of 0.064 K.  The first white contour is 
at 0.51 K.}

\end{figure}

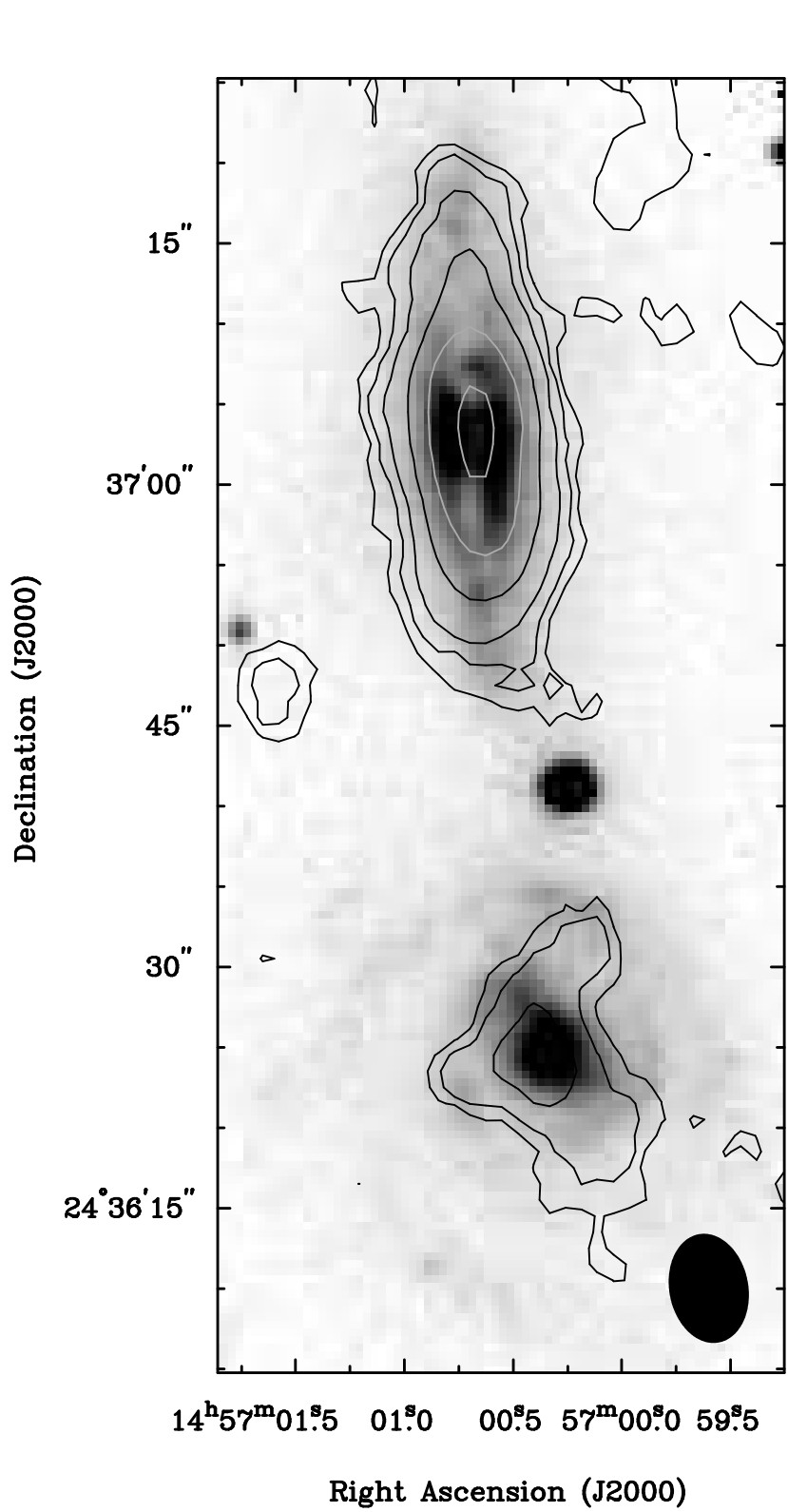
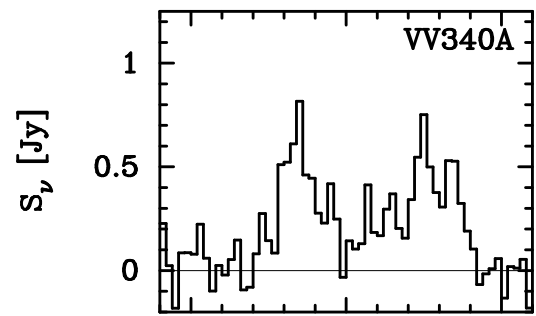
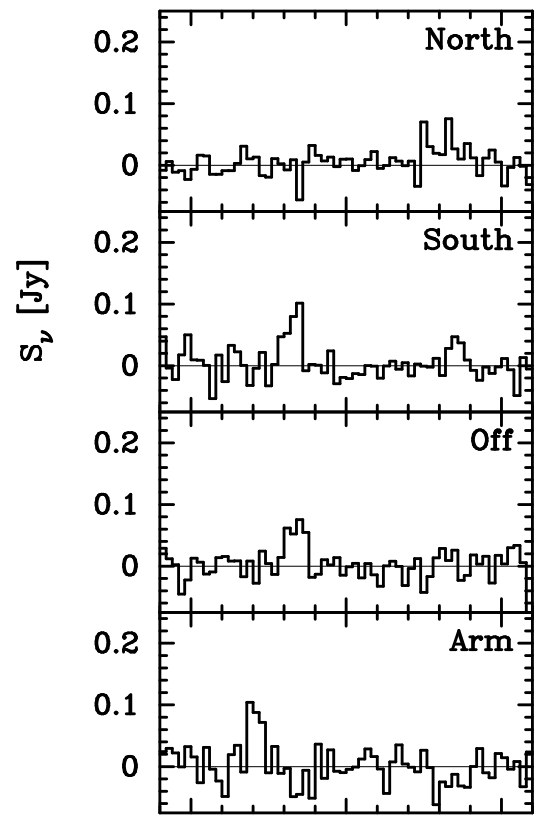
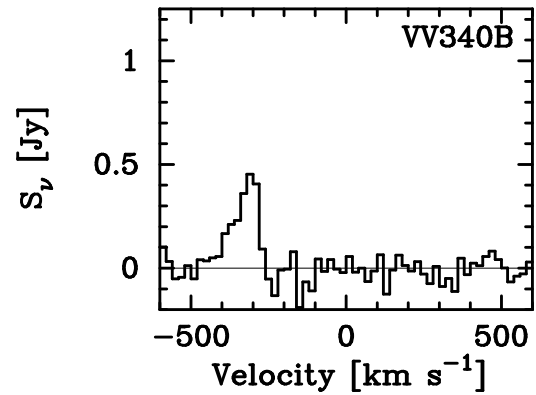
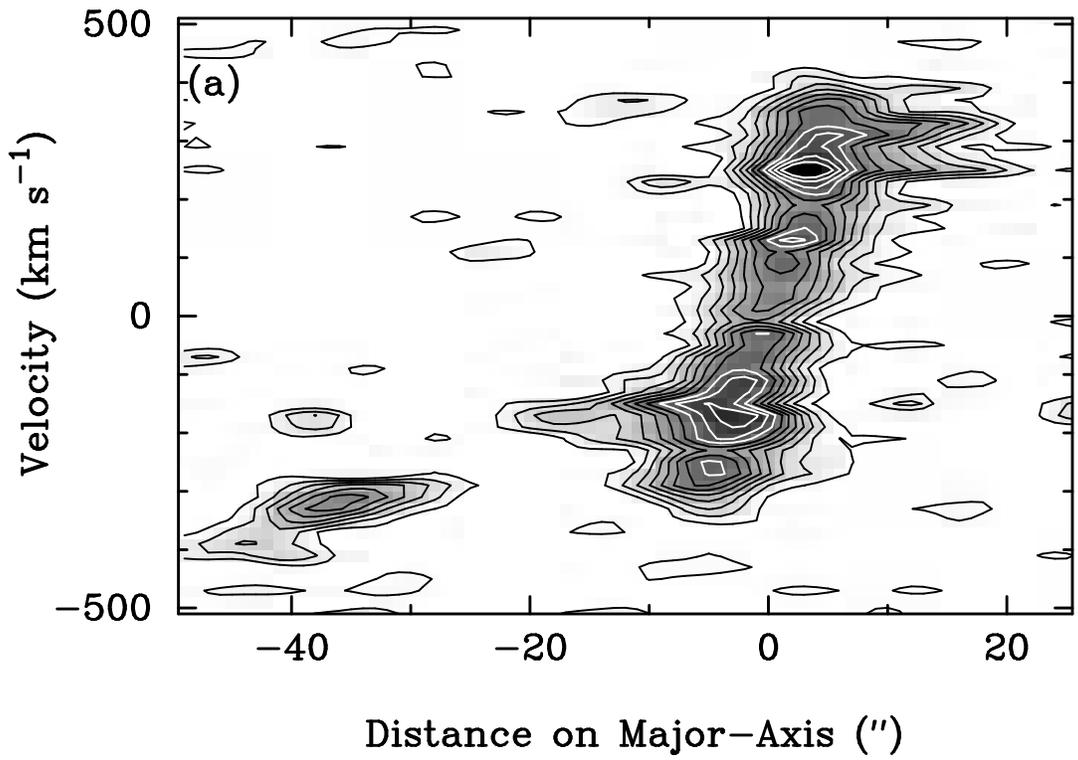
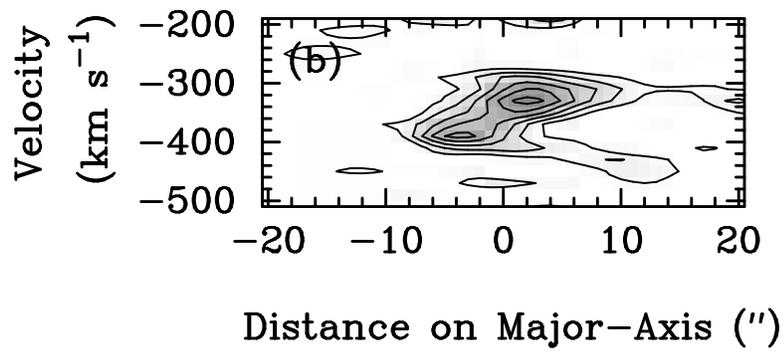
\end{document}